\documentclass[useAMS]{nature}
\usepackage{hyperref} 
\usepackage[sort,numbers]{natbib}
\usepackage{xcolor}
\usepackage[normalem]{ulem}
\usepackage{amsmath}
\usepackage{mathtools}  
\usepackage{graphicx}
\usepackage{gensymb}
\makeatletter
\let\saved@includegraphics\includegraphics
\AtBeginDocument{\let\includegraphics\saved@includegraphics}

\makeatother

\hypersetup{
    colorlinks=true,
    linkcolor=blue,
    filecolor=magenta,      
    urlcolor=cyan,
    citecolor=blue
}
\newcommand{\araa}{Annu. Rev. Astron. Astrophys.}   
\newcommand{\aj}{Astron. J.}   
\newcommand{\apj}{Astrophys. J.}   
\newcommand{\apjl}{Astrophys. J. Lett.}   
\newcommand{\apjs}{Astrophys. J. Suppl. Ser.}   
\newcommand{\ao}{Appl. Opt.}   
\newcommand{\aap}{Astron. Astrophys.}   
\newcommand{\mnras}{Mon. Not. R. Astron. Soc.}   
\newcommand{\pasp}{Publ. Astron. Soc. Pac.}   

\title{The size of the Milky Way galaxy}
\author{Jianhui Lian$^{1*}$, Gail Zasowski$^{2}$, Bingqiu Chen$^{1}$, Julie Imig$^{3}$, Tao Wang$^{1}$, Nicholas Boardman$^{4}$, Xiaowei Liu$^{1}$\\
\small $^{1}${South-Western Institute for Astronomy Research, Yunnan University, Kunming, Yunnan 650091, People's Republic of China}\\
\small $^{2}${Department of Physics \& Astronomy, University of Utah, Salt Lake City, UT 84112, USA}\\
\small $^{3}${Department of Astronomy, New Mexico State University, P.O.Box 30001, MSC 4500, Las Cruces, NM, 88033, USA}\\
\small $^{4}${School of Physics and Astronomy, University of St Andrews, North Haugh, St Andrews KY16 9SS, UK}\\
}

\usepackage{lineno}

\begin{document}

\maketitle

{\bf The size of a galaxy is one of the fundamental parameters that reflects its growth and assembly history. Traditionally, the size of the Milky Way has been characterized by the scale length of the disk, based on the assumption of an exponential density profile. Earlier scale length measurements suggest the Milky Way is an overly compact galaxy, compared to similar galaxies of its mass. These size measurements, however, ignore the presence of the bulge, and the assumption of a single-exponential disk  profile faces growing challenges from the recent observations. The half-light radius is an alternative size measurement that is independent of the galaxy density profile and has been widely used to quantify the size of external galaxies. Here we report the half-light radius of the Milky Way, derived from a new measurement of the age-resolved Galactic surface brightness profile in an unprecedentedly wide radial range from ${\rm R=0}$ to 17~kpc. We find a broken surface brightness profile with a nearly flat distribution between 3.5 and 7.5~kpc, which results in a half-light radius of 5.75$\pm$0.38~kpc, significantly larger than the scale-length inferred from the canonical single-exponential disk profile but in good consistency with local disk galaxies of similar mass. Because our density profile can be decomposed by stellar age and extrapolated backwards in time, we can also confirm that the size history of the Milky Way is broadly consistent with high-redshift galaxies but with systematically smaller size at each look back time. 
Our results suggest that the Milky Way is a typical disk galaxy regarding its size and has likely experienced inefficient secular size growth.}

Our understanding of the Milky Way's structure has improved tremendously through the advancement of Galactic observations over the last decades \citep{vanderKruit2011,bland2016}. Thanks to the proximity to our home galaxy, we are able to study the Milky Way's sub-structures (e.g., disk scaleheights \citep{bovy2012}, spiral arm mapping, and bar/X-shape \citep{shen2020}) in great detail. For the same reason, however, a global picture of Galactic structure is still incomplete. The Sun's position embedded in the disk results in high line-of-sight extinction towards the densest region in the Galaxy, and thus costly observing time to collect data from large samples of stars over a wide spatial range. 

The advent of massive stellar spectroscopic surveys (e.g., APOGEE, LAMOST, Gaia) is now changing the situation rapidly. In particular, the APOGEE survey --- a high-resolution spectroscopic survey in the near-infrared --- has mapped $\sim0.6$ million stars over an unprecedentedly wide range in Galactocentric radius and disk height \citep{zasowski2013,majewski2017,zasowski2017,beaton2021,santana2021}. Operating in the near-infrared enables detection of stars in the disk plane, in particular in the Galactic center direction, where the extinction is high, and thus a unique sampling of stars with full radial coverage. This survey provides an ideal dataset to study the structure of the Milky Way on large scales.  

One of the fundamental parameters of a galaxy's structure is its size. For a long time, the Milky Way has been treated as a canonical disk galaxy with single-exponential profile, and the size characterized by the scale length of the exponential component \citep{juric2008,bovy2012,bland2016}. This scale length is noticeably ``short'' when compared to large populations of disk galaxies of the same mass and luminosity \citep{vanderKruit2011,licquia2015,boardman2020_MWA}. However, many recent Galactic structure studies based on massive stellar spectroscopic surveys reveal a complex density distribution of the disk, with possibly more than one exponential component \citep{bovy2016,mackereth2017,lian2022b}. In light of this and the presence of the significant bar/bulge, an alternative size quantity --- independent of the assumed density distribution --- is needed to better describe the global structure of the Milky Way. 

Half-light (i.e., effective) radius is just such a size quantity that is widely used in extragalactic studies \citep{shen2003,blanton2011}. The half-light radius, ${\rm R_{\rm 50}}$, is defined as the radius within which half of a galaxy's luminosity is emitted. In this work, we present the Milky Way's half-light radius derived from the luminosity surface density profile based on APOGEE data, and perform a direct comparison with other similar-mass, star-forming disk galaxies in both the local and high-redshift Universe. 

The key to determine the half-light radius is to derive the surface brightness profile over a full radial range. While for external galaxies it is often straightforward to extract this profile from images, it has long been a challenge for the Milky Way, especially at small radii, because of, for example, high extinction in the disk and strong selection effects and incomplete spatial sampling in observations. As a result, few attempts has been made before to derive the surface brightness profile of the Milky Way and the profiles obtained so far are all beyond the solar radius at ${\rm R}>8$~kpc \citep{xiang2018}. In this work, we take advantage of the wide spatial coverage of the APOGEE survey and derive the non-parametric surface brightness profile of the Milky Way in an unprecedented radial range from ${\rm R=0}$ to 17~kpc, for the first time enabling a direct measurement of the Milky Way's half-light radius. We use chemical abundances, ages, and distances of individual stars observed with APOGEE \citep{majewski2017} and Gaia \citep{gaia2016,gaia2021}. We transform the {\it observed} density of stars to the {\it intrinsic} density of the underlying stellar population by correcting for the survey selection function, separately for stars of different abundances. The 3D luminosity density distributions of the intrinsic populations of different abundances are then integrated in the vertical direction to obtain the luminosity surface density profile of mono-age and entire stellar populations (Methods). 

When including stars of all ages from 0 to 12~Gyr, the luminosity surface density profile is rather complex (Figure~1). The inner upturn at ${\rm R<3.5}$~kpc is driven by the Galactic bulge \citep{rojas2019}. 
The surface density profile of the disk consists of multiple components: a nearly flat plateau between 3.5 and 7.5~kpc, an exponential component between 7.5 and 14~kpc with scale length consistent with previous studies of the structure of the Galactic disk \citep{bland2016}, and a steep decrease beyond 15~kpc. Such a broken disk density profile clearly deviates from the widely-assumed, canonical single-exponential profile of the Galactic disk and will largely reshape our understanding of the Milky Way's structure. The flattened inner disk density profile is in broad agreement with recent studies on the structure of mono-abundance populations based on different spectroscopic surveys \citep{lian2022b}. We further validate this feature using Gaia data with robust parallax observations (Methods). Such a down-bending disk profile with inner part being flatter is frequently seen in local disk galaxies \citep{pohlen2006,erwin2008}, but the break usually occurs in low-density environment at the galaxy peripheries. Some rarer cases have the profile breaks in the inner disk (Methods), suspected to originate from stellar-dynamical side effects of the bar-formation process \citep{erwin2008}.  

With the luminosity surface density profile, we for the first time measure the half-light radius of the Milky Way and obtain a result of 5.75$\pm$0.38~kpc. This measurement does not strongly depend on the choice of the wavelength of the monochromatic luminosity or the bolometric luminosity for the density profile (Methods). Because the inner disk profile flattens, the half-light radius of the Milky Way is significantly larger than that expected from a canonical picture of the Milky Way's structure with a bulge and single-exponential disk components ($3.43\pm0.43$~kpc would be measured under this assumption, Methods). 

When dividing into mono-age populations, the surface density profile varies systematically with stellar age. In the bulge region (${\rm R}<3.5$~kpc), except for the oldest age bin (10-12 Gyr), younger populations have generally lower luminosity density, which is broadly consistent with the bulge star formation history \citep{zoccali2003,nataf2016,hasselquist2020,lian2020c}. The mono-age population of 8-10 Gyr exhibits the highest excess over the disk, which might indicate the bar formation in the Galactic center \citep{bovy2019}. In the disk region (${\rm R}>3.5$~kpc), younger populations show steeper positive slopes within $7.5$~kpc and flatter slopes beyond it, with a consistently steep negative slope at $\rm R>7.5$~kpc for the youngest age bin of 0-2 Gyr.The flat inner disk profile of the summed populations is a combination of the negative slopes of old populations and then positive slopes of young populations in this region. Given the increasing slope with age, especially considering the old populations have experienced the most radial migration, the inner disk profile of the Milky Way as a whole is steeper at earlier look back time. In the outer disk beyond 15~kpc, the steep decrease of density profile might indicate the truncation of the Galactic disk \citep{minniti2011}, beyond which the stars are populated possibly via radial migration from inner regions \citep{martinez2009,ruiz-lara2017,lian2022a}. The steepest slope of the youngest population is consistent with the radial migration origin of the outer disk as younger populations are expected to experience less migration. 

To compare the Milky Way with other galaxies, we select a sample of 19092 low-redshift ($0 < z < 0.1$), massive (${\rm log(M_*})>10{\rm M_{\odot}}$), face-on star-forming disk galaxies from SDSS \citep{york2000}. The half-light radii of these galaxies have been measured based on the SDSS images and provided in the NASA-Sloan Atlas (NSA) catalog \citep{blanton2011}. The left panel of Figure~2 shows the position of the Milky Way in the mass-size plane in comparison with the low-redshift disk galaxies. For the Milky Way's total stellar mass, we adopt the stellar mass of the disk \citep{xiang2018} and that of the bulge excess over the disk \citep{licquia2015} from the literature, which in combination results in a total stellar mass of $4.81\pm0.13\times10^{10}{\rm M_{\odot}}$ (Methods). We have tested that adopting a lower stellar mass of the Milky Way to account for the flattened inner disk profile does not change our results qualitatively. For reference, we also calculate the half-light radius based on a canonical picture of Milky Way's structure that consists of a bulge and a single-exponential disk with scale length of 2.6~kpc \citep{bland2016} (Methods). The obtained half-light radius is 3.43$\pm$0.43~kpc, a factor of 1.7 smaller than that measured from the non-parameteric surface brightness profile in Fig.~1. 

A direct comparison between the size of the Milky Way and other galaxies with the same stellar mass (i.e., within the gray vertical band in the left panel) is shown in the right panel of Figure~2. Comparing to the Milky Way-mass disk galaxies, the half-light radius of the Milky Way based on the canonical picture of Galactic structure is extraordinary small, lower than 95\% galaxies in the comparison sample and putting the Milky Way among the most compact galaxies. In contrast, the half-light radius measured from the non-parametric surface density profile is in good consistency with other disk galaxies, suggesting the Milky Way is not compact but a typical disk galaxy regarding its size. Interestingly, this large half-light radius measurement leads to a steep integrated stellar and gas metallicity gradient normalized to the effective radius \citep{boardman2020_MWA,lian2023}, exacerbating the rarity of the Milky Way among local galaxies regarding the metallicity gradient.  

The temporally-resolved observations in the Milky Way enable an intriguing inspection of the size growth history of the Galaxy and comparison with observations of high-redshift galaxies. In Figure~3, we present the measurement of half-light radius of mono-age populations (blue) and the summed populations of the Milky Way (orange) at different look back times. The growing size of mono-age populations supports inside-out formation of the Milky Way \citep{bovy2012}. The lack of growth, or even shrinking size, of the youngest population may point to the end of this inside-out formation process. 

The half-light radius of high-redshift galaxies are estimated using the empirical size-mass-redshift relation of disk galaxies from redshift of 0.25 to 3 \citep{vandelWel2014}. 
We estimate the stellar mass of the Milky Way at different look back time 
and apply them to the size-mass-redshift relation in \citep{vandelWel2014} to obtain the size of the same mass galaxies at corresponding redshifts (Methods). The size growth history of the Milky Way is broadly consistent in its slope (kpc/Gyr) with that of high-redshift disk galaxies on average, but it has systematically smaller size at each look back time.   The significance of the discrepancy in size between the Milky Way and high-redshift galaxies is less than it appears as the uncertainties in the size estimate of high-redshift galaxies represent the error of the mean instead of the standard deviation of the distribution. Nevertheless, the systematically smaller size suggests inefficient secular size growth of our Galaxy from the high-redshift Universe to the present day, possibly a sign of limited radial migration. 

To summarise, we find a rather complex luminosity surface density profile of the Milky Way, with a broken disk density profile that flattens between 3.5 and 7.5~kpc. The presence of the bulge and the broken disk profile highlight the necessity to use a non-parametrically-derived half-light radius instead of scale length to characterize the size of the Milky Way. Owing to the flattened inner disk density profile, we obtain a significantly larger size of the Milky Way than that expected from a single-exponential disk profile, pushing the Milky Way from an outlier in the galaxy population to a typical one regarding the size. For mono-age populations, their half-light radii increase monotonically from early times until an age of 2~Gyr, consistent with an inside-out formation of the Milky Way that possibly ends in the last few Gyrs. Comparing to high-redshift galaxies, the Milky Way has experienced a similar size growth history but with systematically smaller size, which indicates inefficient secular size growth of our Galaxy.  

{\bf Method}

{\bf Data.} This work is based on the last data release of the APOGEE survey within the SDSS-IV Data Release 17 (DR17) \citep{blanton2017}. APOGEE is a near-infrared, high-resolution spectroscopic survey \citep{majewski2017} that provides high-quality measurements of stellar parameters (e.g., log(g) and T$_{\rm eff}$) and elemental abundances (e.g., [Fe/H], [Mg/Fe]) for $\sim0.6$ million stars in nearly 1000 discrete fields that are semi-regularly distributed throughout the Galactic disc, bulge, and halo \citep{zasowski2013,zasowski2017,beaton2021,santana2021}. The target sample used in this work are those from the Main Red Star Sample (see a description of the flags used to identify the main sample in \url{https://www.sdss.org/dr16/algorithms/bitmasks/}).
These stars are randomly selected, on a field-to-field basis, from candidates defined in the 2MASS $H$--$(J-Ks)_0$ colour-magnitude diagram. The spectra of these stars are obtained using custom spectrographs \citep{wilson2019} with the 2.5~m Sloan Telescope at the Apache Point Observatory and with the 2.5~m Ir\'en\'ee du~Pont telescope at Las Campanas Observatory \citep{bowen1973,gunn2006}. 
The spectra are reduced and elemental abundances and stellar parameters of individual stars are produced by custom pipelines using a custom line list (ASPCAP) \citep{nidever2015,garcia2016,smith2021}.
The stellar ages and spectro-photometric distances are derived by applying the astroNN deep-learning code to the spectroscopic data from APOGEE, asteroseismic data from APOKASC \citep{pinsonneault2018,mackereth2019}, and astrometric data from Gaia, and are provided in the astroNN Value Added Catalog \url{https://data.sdss.org/sas/dr17/env/APOGEE_ASTRO_NN} with typical age uncertainties of 30\% and distance uncertainties of 10\% \citep{mackereth2019,leung2019}. Throughout this paper, we assume the cosmological parameters ($\Omega_{\rm M},\ \Omega_{\Lambda}, h_0$) = (0.27, 0.73, 0.71). 

{\bf The luminosity surface density profile and half-light radius.} Our luminosity surface density profiles are derived using the 3D density distributions of mono-abundance populations (MAPs) after carefully correcting for the APOGEE survey selection function \citep{lian2022b}. The procedure is described in detail in Lian et al.\, \citep{lian2022b,lian2023}, but can be summarized in four steps:
\begin{enumerate}
\item Calculate the probability that a star, at a given 3D Galactic position and with given [Fe/H] and [Mg/Fe] abundances, would be observed by the APOGEE survey based on the Padova stellar evolution models \citep{bressan2012,chen2014,chen2015,paola2017} and 3D dust extinction map \citep{bovy2016_dust}; 
\item Divide the observed number density of APOGEE stars at this position and abundance by the observational probability to obtain the intrinsic number density; 
\item Sample the stellar isochrones \citep[assuming a Kroupa IMF;][]{kroupa2001} to derive the 3D luminosity density distribution of the underlying MAPs;
\item Unfold each MAP at a given position along the age dimension assuming the observed age distribution of the MAP at the same position to obtain the luminosity density distribution of mono-age populations from 0 to 12~Gyr, with bin width of 2~Gyr; 
\item At each radial bin from 0 to 17~kpc, with even step size of 1~kpc, fit the vertical density distribution of each separate MAP with an exponential function and integrate the best-fitted density model to obtain the luminosity surface density for that MAP. 
\end{enumerate}
We consider MAPs with [Fe/H] from $-0.9$ to +0.5, defined by a bin width of 0.2~dex, and [Mg/Fe] from $-0.1$ to 0.4, with a bin width of 0.1~dex. 


To measure the half-light radius, we first obtain the luminosity surface density profile of the entire population by summing up those of individual mono-age populations. Then, we linearly interpolate the luminosity surface density profile onto a finer grid with step of 0.01~kpc. Finally, we calculate the cumulative luminosity profile to identify the radius that contains half of the total luminosity (i.e. half-light radius, ${\rm R_{50}}$). The half-light radius at different look back times is derived from the luminosity surface density profile of mono-age populations following the same procedure, excluding the populations younger than the given lookback time.


{\bf Uncertainties of the half-light radius measurement.} The uncertainties of the half-light radius measurement presented in this work are inherited from uncertainties in the luminosity surface density profiles, which are contributed by the luminosity surface density measurement and the average radial position of the radial bins. For each radial bin, the uncertainty of the radial coordinate is assumed to be the standard deviation of the radii of individual 3D density measurements within this radial bin. The uncertainties of the surface density measurements originate from the APOGEE selection function and the observed number density of APOGEE stars at each 3D position. While Poisson error is assumed for the latter, it is difficult to directly estimate the uncertainties of the selection function, which are dominated by the systematic uncertainties of the 3D extinction map and the stellar evolution models. Thus we take an empirical approach to estimate these uncertainties, as described below.

We first estimate the stochastic uncertainty of the half-light radius using a Monte Carlo simulation that considers the uncertainties of the observed 3D number densities and radial positions of each luminosity surface density measurement. That is, we measure $R_{50}$ using the method above but varying each observed 3D number density and radial coordinate within their uncertainties, re-computing the luminosity surface density profiles, in a series of 100 trials. The standard deviation of derived $R_{50}$ values, taken as the stochastic uncertainty, is 0.14~kpc. To obtain an approximate understanding of the systematic uncertainties of the half-light radius measurement originating from the selection function correction, we adopt a different stellar evolution model and a different 3D extinction map for the bulge (the most uncertain region of the map) and repeat the whole analysis above to derive the half-light radius. Using the MESA isochrones \citep{choi2016,dotter2016}, a half-light radius of 6.03~kpc is obtained, 0.28~kpc longer than the measurement derived using the Padova isochrones. Using the extinction map of the bulge (i.e., Galactic longitude $|l|<10\degree$ and latitude $|b|<5\degree$) from \citep{chen2013}, a higher luminosity surface density of the bulge is obtained, which results in a half-light radius of 5.54~kpc, 0.21~kpc shorter than the original measurement. 
In light of these tests, in this work, to be conservative we assume a systematic uncertainty of 0.35~kpc (i.e., $\sqrt{0.28^2+0.21^2}$) for our half-light radius measurement, which in combination with the stochastic error gives an total uncertainty of 0.38~kpc.   

The center of the Milky Way hosts a nuclear stellar disc with size of a few hundred parsecs \citep{launhardt2002}. This structure is unlikely to be well represented in the APOGEE survey due to its compact size. To test the effect of including this structure in the estimate of half-light radius of the Milky Way, we manually increase the luminosity surface density at the innermost radial bin of 0-1~kpc by the amount of mass of the nuclear stellar disc \citep{sormani2020}. As a result, a slightly shorter half-light radius of 5.67~kpc is obtained,  which is consistent with the measurement in the main paper within 1~$\sigma$. 

{\bf The Milky Way's total stellar mass.} The size comparison between the Milky Way and galaxies in the local and high-redshift universe depends on the adopted value of the Milky Way's stellar mass. 

For the size comparison, in the main paper we estimate the Milky Way's total stellar mass based on recent reports of the stellar mass of Milky Way's disk and bulge from the literature. 
For the disk component, we take the measurement of 3.6$\pm0.10\times10^{10}\ {\rm M_{\odot}}$ from \citep{xiang2018}, which takes a careful treatment of the selection effect of the LAMOST survey. \cite{xiang2018} estimate the disk stellar mass by integrating over a single exponential profile best-fitted to the surface density distribution at R$>8$~kpc and extrapolated to the Galactic center (but see below). This measurement is based on the assumption of 8.0~kpc for the solar radius and a scale length measurement of 2.48~kpc. To be consistent with this work where a solar radius at 8.2~kpc is assumed, a conversion factor of $1.08$ is applied. For the bulge component, we adopt the mass of 0.91$\pm0.07\times10^{10}\ {\rm M_{\odot}}$ reported in \citep{licquia2015}, comprising a compilation and Bayesian analysis of previous measurements. Combining the disk and bulge together, a total stellar mass of $4.81\pm0.13\times10^{10}{\rm M_{\odot}}$ is obtained for the Milky Way as a whole.  

The flattened inner disk density profile reported here would lead to a lower estimate of the disk stellar mass than that assuming single exponential profile, i.e., in \cite{xiang2018}. Here we perform a test to understand how the size comparison would be affected by this effect. A more detailed study on the estimate of Milky Way's stellar mass affected by the flattened distribution is beyond the scope of this work and will be included in a future work. 

The disk mass measurement in \citep{xiang2018} relies on a single exponential profile fitted to the data at $R>8$~kpc. If assuming flattened distribution in the inner Galaxy at $R<7.5$~kpc, we obtain a considerably lower estimate of the disk stellar mass by a factor of 1.7 (Lian et al. in prep.). Assuming the bulge mass from \citep{licquia2015}, this results in a Milky Way's total stellar mass of $3.20\pm0.13\times10^{10}{\rm M_{\odot}}$. With this estimate, the half-light radius of the comparison sample in the right panel of Fig.~2 would shift left towards smaller values by $\sim$1~kpc, with an average value of $\sim$5.6~kpc, because of the positive mass-size relation. In this case, the half-light radius of the Milky Way reported in this work is even closer to the average value of other disk galaxies, while the estimate based on a single exponential disk is still significantly smaller than the average of the sample. Therefore, our conclusion that the Milky Way is a normal galaxy regarding its size is still valid after considering the flattened disk profile in the stellar mass estimate of the Milky Way. 

{\bf Verification of radial density profile based on Gaia.} To verify the flattened inner disk density profile identified in Fig.~1, we conduct an independent investigation of the radial density distribution using observations from Gaia early Data Release 3 \citep{gaia2016,gaia2021} with good parallax measurements, which provide independent robust distance estimates. To focus on the radial direction, for simplicity we select Gaia stars strictly along the Galactocentric (Galactic longitude $|l|<4\degree$ and Galactic latitude $|b|<0.5\degree$) and the anti-Galactocentric ($176<l<184\degree$, $|b|<0.5\degree$) directions. 
At a distance of 2~kpc, Gaia parallax observations are complete at $G$-band magnitude down to 17~mag. Therefore we further restrict the sample with Gaia $G$-band magnitude within [11,17]~mag and distance within 2~kpc. Finally, 55962 and 21278 stars remain in the Galactocentric and anti-Galactocentric samples. The spatial coverage of the two samples in the X-Y plane is illustrated in the insert sub-panel in the upper left panel of Figure~4. 

To correct for the selection effect, we follow a similar process to what we have done for the APOGEE sample. At each distance, we simulate the apparent $G$-band magnitude distribution and estimate the fraction of mock stars falls into the magnitude cut adopted in our sample selection. To this purpose, for simplicity, we sampling the PARSEC isochrone \citep{bressan2012} with solar-age and solar-metallicity assuming Kroupa IMF and apply the average extinction within the Galactocentric and anti-Galactocentric directions estimated from the 3D extinction map \citep{bovy2016_dust}. Then we divide the observed number density of stars at each distance (approximately equivalent to radius) by this fraction which gives rise to the intrinsic number density of the underlying population. The obtained 3D density profile along the radial direction on the disk plane is shown in the right panel of Figure~4. 

The obtained density profile based on Gaia data alone also present a clear flattened density distribution in the inner Galaxy, with the outer part being well explained by the canonical exponential disk with frequently reported scale length of 2.6~kpc. This flattened density distribution in the inner Galaxy is unlikely to 
disappear when converting the 3D density profile into 1D surface density profile given the limited variation of disc thickness over the radial range of the flat distribution. Therefore Gaia data alone provide independent evidence for the existence of flattened inner disk profile. 

A caveat for the readers is that this is a rough estimate of the radial 3D density distribution based on Gaia data. A number of assumptions are made to simplify the calculation, such as the selection effect is estimated based on a single stellar population in age and abundance and a smooth dust extinction is assumed within the Galactocentric and anti-Galactocentric directions. These simplifications possibly cause the up-and-down in the inner Galaxy density distribution. A more careful study would need to consider the rich variety of stellar populations in age and abundances and the potential variation of dust extinction, in particular in the Galactocentric direction, which are beyond the scope of this paper.

{\bf Example external galaxy with flattened inner disk profile.} The flattened inner disk profile reported in this work is not unique in external galaxies \citep{pohlen2006,erwin2008}. Upper right panel of Figure~4 shows one of the galaxies with such flattened profile in the inner part. It is a massive (${\rm M_*=4.32\times10^{10}M_{\odot}}$), star-forming disk galaxy 
at redshift of 0.062 observed in SDSS survey. The surface brightness profile of this galaxy consists of an inner upturn within 3~kpc, a flattened profile between 3 and 6~kpc, and an exponential disk beyond 7~kpc, resembling the luminosity surface density profile of the Milky Way reported here.   

{\bf Surface density profile of monochrome luminosity in the optical and near-infrared.} To test whether our measurement of the half-light radius is dependent on the wavelength of the luminosity, we derive the luminosity surface density profile of the monochrome luminosities in the optical r-band and near-infrared H-band in contrast to the the bolometric luminosity presented in the main paper. The obtained monochrome luminosity density profiles are rather similar to that of the bolometric luminosity (bottom left panel of Figure~4), indicative of minor variation of the Milky Way's structure measured in different wavelength. The half-light radius estimated from these surface density profiles of monochrome luminosity are indeed comparable, with that in the H-band being mildly smaller than the one in the r-band by $0.45$~kpc. 

{\bf Parametric luminosity surface density profile and half-light radius} To estimate the half-light radius based on the canonical picture of the Milky Way's structure, i.e. bulge and single exponential disk, we generate a simple parametric luminosity surface density profile as shown in the bottom right panel of Figure~4. For the single exponential disk component, we adopt a scale length of 2.6~kpc \citep{bland2016} and solar surface mass density of 35.3~${\rm M_{\odot}pc^{-2}}$ from \citep{xiang2018} and mass-to-light ratio of 0.76 \citep{lian2022b}. For the bulge component, we assume the same shape as the excess over the disk at ${\rm R<3.5}$~kpc in the luminosity surface density profile in Fig.~1, which can be approximately described by a steep exponential profile with scale length of 0.56~kpc.  
Since the half-light radius is larger than the region dominated by the bulge, the assumed shape of the bulge density profile does not strongly affect the estimate of the half-light radius. 
We adjust the amplitude of the bulge density profile to reach a total luminosity of 1.12$\times10^{10}{\rm L_{\odot}}$ within ${\rm R=3.5}$~kpc, converted from the total bulge mass of 0.91$\times10^{10}{\rm M_{\odot}}$ \citep{licquia2015} assuming mass-to-light ratio of 0.81 \citep{lian2022b}. The half-light radius derived for this parametric luminosity surface density profile is 3.43$\pm0.43$kpc, significantly smaller than that of the Milky Way when taking the flattened inner disk profile into account and of external galaxies (Fig.~2). 

{\bf Data Availability.} All data presented in this work are available in public repository \url{https://github.com/lianjianhui/Source-data-for-MW-size-paper.git}.

{\bf Acknowledgement.} 
Funding for the Sloan Digital Sky Survey IV has been provided by the Alfred P. Sloan Foundation, the U.S. Department of Energy Office of Science, and the Participating Institutions. SDSS-IV acknowledges
support and resources from the Center for High-Performance Computing at the University of Utah. The SDSS web site is www.sdss.org.

SDSS-IV is managed by the Astrophysical Research Consortium for the 
Participating Institutions of the SDSS Collaboration including the 
Brazilian Participation Group, the Carnegie Institution for Science, 
Carnegie Mellon University, the Chilean Participation Group, the French Participation Group, Harvard-Smithsonian Center for Astrophysics, 
Instituto de Astrof\'isica de Canarias, The Johns Hopkins University, Kavli Institute for the Physics and Mathematics of the Universe (IPMU) / 
University of Tokyo, the Korean Participation Group, Lawrence Berkeley National Laboratory, 
Leibniz Institut f\"ur Astrophysik Potsdam (AIP),  
Max-Planck-Institut f\"ur Astronomie (MPIA Heidelberg), 
Max-Planck-Institut f\"ur Astrophysik (MPA Garching), 
Max-Planck-Institut f\"ur Extraterrestrische Physik (MPE), 
National Astronomical Observatories of China, New Mexico State University, 
New York University, University of Notre Dame, 
Observat\'ario Nacional / MCTI, The Ohio State University, 
Pennsylvania State University, Shanghai Astronomical Observatory, 
United Kingdom Participation Group,
Universidad Nacional Aut\'onoma de M\'exico, University of Arizona, 
University of Colorado Boulder, University of Oxford, University of Portsmouth, 
University of Utah, University of Virginia, University of Washington, University of Wisconsin, 
Vanderbilt University, and Yale University.

\begin{figure*}
	\centering	\includegraphics[width=12cm]{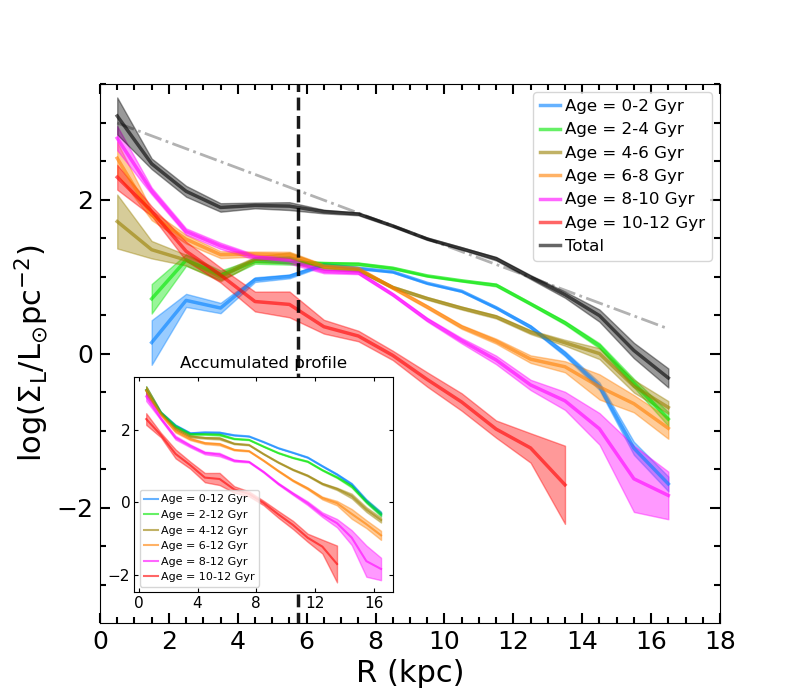}
	\caption{The luminosity surface density profiles of the integrated (black) and mono-age (colourful) stellar populations of the Milky Way. Shaded regions indicate the 1$\sigma$ uncertainty on the luminosity surface densities, and the vertical dashed line marks the half-light radius of 5.75~kpc. The grey dash-dotted line illustrates a single exponential profile with a scale length of 2.6~kpc\citep{bland2016}, normalized to the surface density of the radial bin between 8 and 9~kpc. The inset panel shows the accumulated luminosity surface density profile as a function of age.          
	} 
	\label{surf-age}
\end{figure*} 

\begin{figure*}
	\centering	\includegraphics[width=16cm]{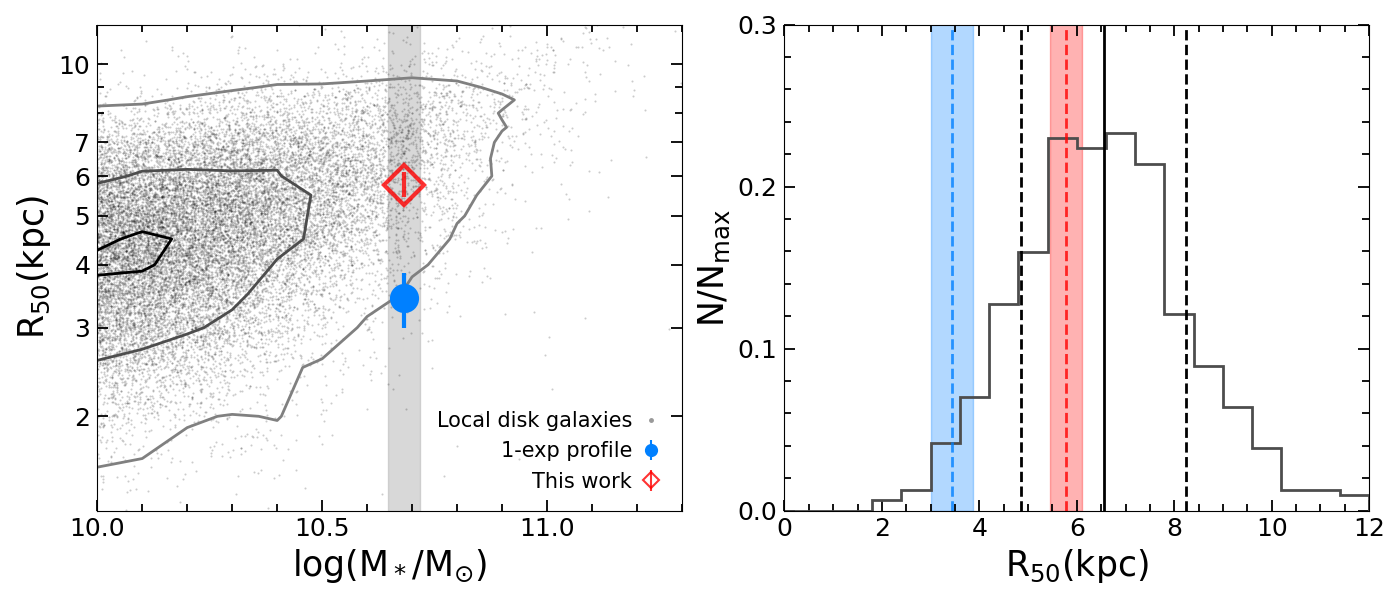}
	\caption{The half-light radius of the Milky Way in comparison with that of local disk galaxies. {\sl Left}: The position of the Milky Way in the mass-size plane of galaxies. Two half-light radius measurements of the Milky Way are presented, one derived from the non-parametric surface brightness profile shown in Fig~1 (red diamond) and the other derived assuming a canonical bulge and single exponential disk (blue circle). The grey vertical stripe indicates the stellar mass range used to select Milky Way-mass disk galaxies shown in the right panel. {\sl Right:} Comparison between the half-light radius of the Milky Way with Milky Way-mass disk galaxies in the local Universe. Red and blue dashed lines and shaded regions represent the two half-light radius measurement of the Milky Way shown in the left panel and their $1\sigma$ uncertainties. Solid and dashed black lines denote the mean half-light radius of the comparison sample and the 1$\sigma$ scatter, respectively.      
	} 
	\label{mass-size}
\end{figure*} 

\begin{figure*}
	\centering	\includegraphics[width=12cm]{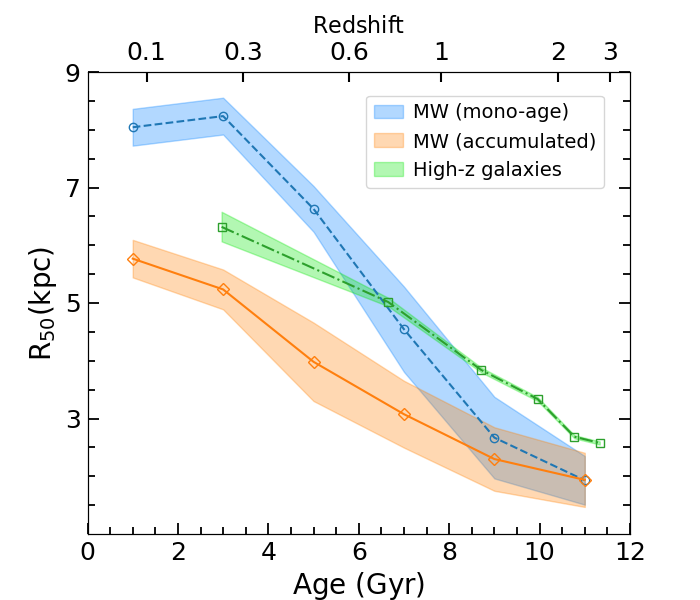}
	\caption{The time evolution of the Milky Way's half-light radius.
 Blue circles and shaded region represent the half-light radii and 1$\sigma$ uncertainties of mono-age populations. Orange diamonds and shaded region indicate the half-light radius and 1$\sigma$ uncertainty of the summed galactic population at each look back time. Green squares and shaded region are the half-light radius and 1$\sigma$ uncertainty of higher redshift disk galaxies with Milky Way's stellar mass at the corresponding look back time, calculated from the relationships in \citep{vandelWel2014}.     
	} 
	\label{radius-age}
\end{figure*} 

\begin{figure*}
	\centering	\includegraphics[width=14cm,viewport=20 50 900 810,clip]{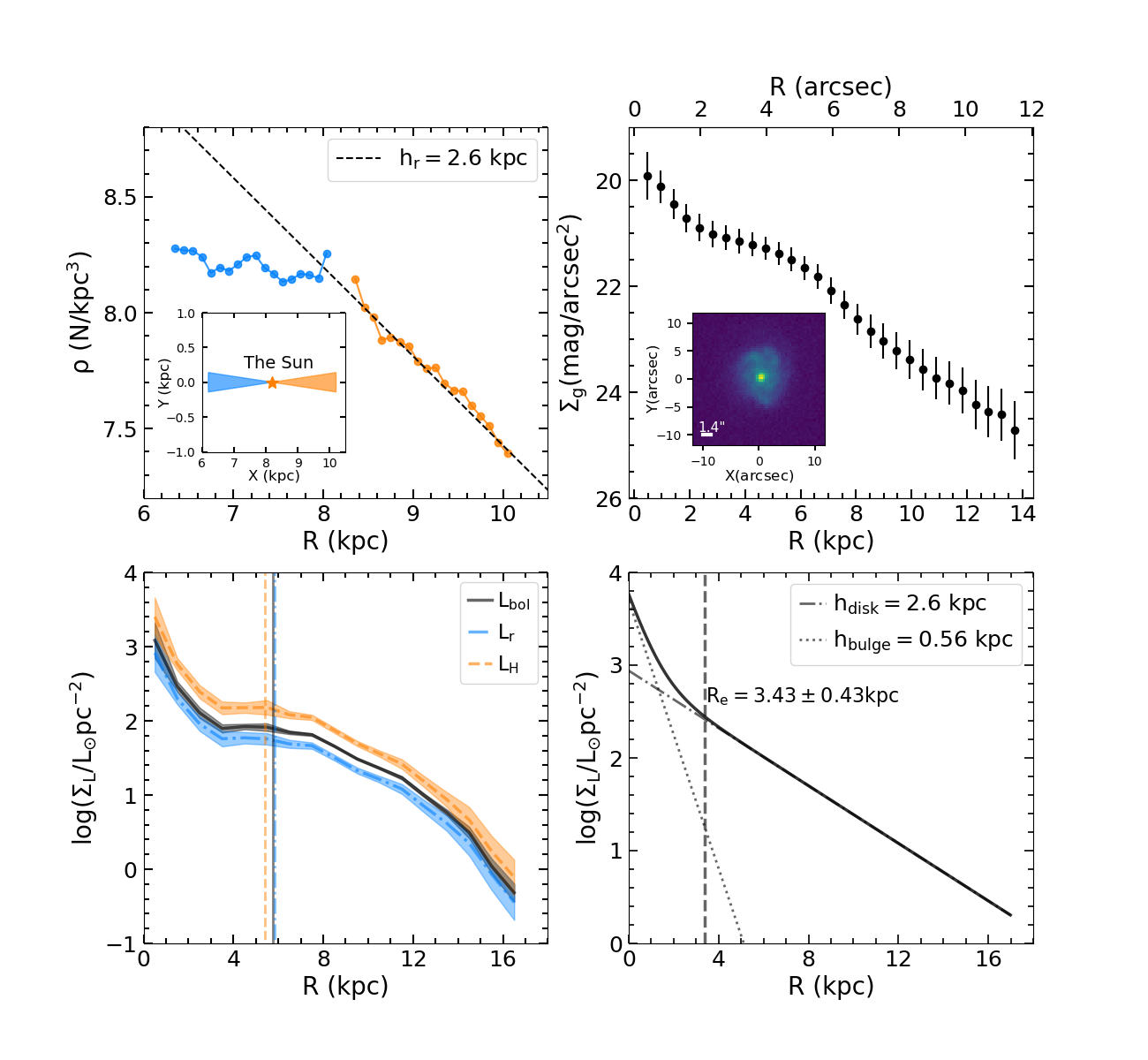}
\caption{Luminosity radial density profiles of the Milky Way and an example external galaxy. {\sl Upper left: } The 3D number density distribution along the radial direction on the disk plane based on Gaia data that have been corrected for the selection function. The dashed line indicate a single exponential profile with a scale length of 2.6~kpc. The insert panel illustrates the spatial coverage of our Gaia Galactocentric (blue) and anti-Galactocentric (orange) samples in the X-Y plane. {\sl Upper right: } An example galaxy with a flat density distribution at intermediate radial range, from SDSS. The surface brightness profile in optical $g$-band is extracted from the image shown in the insert panel. {\sl Bottom left: } The non-parametric surface density profile of the Milky Way in bolometric luminosity and luminosity of single optical r-band and near-infrared H-band. Vertical dotted lines indicate the half-light radius of each profile. {\sl Bottom right: } A parametric luminosity surface profile and associated half-light radius of the Milky Way based on the canonical picture of the Galactic structure that consists of a bulge and an single exponential disk.} 
	\label{gaia}
\end{figure*}

\end{document}